# WHOSE – A Tool for Whole-Session Analysis in IIR

Daniel Hienert[1], Wilko van Hoek[1], Alina Weber[1] and Dagmar Kern[1]

[1] GESIS – Leibniz Institute for the Social Sciences, Cologne, Germany
`firstname.lastname@gesis.org`

**Abstract.** One of the main challenges in Interactive Information Retrieval (IIR) evaluation is the development and application of re-usable tools that allow researchers to analyze search behavior of real users in different environments and different domains, but with comparable results. Furthermore, IIR recently focuses more on the analysis of whole sessions, which includes all user interactions that are carried out within a session but also across several sessions by the same user. Some frameworks have already been proposed for the evaluation of controlled experiments in IIR, but yet no framework is available for interactive evaluation of search behavior from real-world information retrieval (IR) systems with real users. In this paper we present a framework for whole-session evaluation that can also utilize these uncontrolled data sets. The logging component can easily be integrated into real-world IR systems for generating and analyzing new log data. Furthermore, due to a supplementary mapping it is also possible to analyze existing log data. For every IR system different actions and filters can be defined. This allows system operators and researchers to use the framework for the analysis of user search behavior in their IR systems and to compare it with others. Using a graphical user interface they have the possibility to interactively explore the data set from a broad overview down to individual sessions.

## 1 Introduction

Kelly et al. [12] summarize the challenges and problems that arise in the evaluation of Interactive Information Retrieval (IIR) systems. One main goal should be the development of re-usable tools that enable researchers from different domains to investigate search behavior of real users in different environments and produce comparable results. Initial work on this task has been done and frameworks and toolkits have been proposed that allow controlled experiments in different settings [3, 5, 9]. This means that with the help of these frameworks researchers can design, create and conduct laboratory experiments for different domains, different data sets and carefully chosen user groups. Our aim in this work is to extend these set of tools with a tool that (1) supports the analysis of controlled and uncontrolled data sets from real-world IR systems and therefore from real users, (2) can either use existing log files or newly rec-orded data, (3) is based on whole-sessions and multiple sessions and (4) supports the overall process from logging over processing to interactive analysis.

The topic of whole-session evaluation has been recently discussed in a seminar on "Whole Session Evaluation of Interactive Information Retrieval Systems"[1] which has been conducted by members of the IIR community. The main claim of the workshop output is that IR research has concentrated so far on how well an IR system responds to a single query, for example, by presenting a well-ranked result list. However, user interaction in an IR system takes place in the context of a search session. A session is not limited to a single query and some matching documents, but comprises all interactions, queries, resulting documents as well as the user′s learning process about the topic and the system.

In this paper we present an analysis tool for whole-session analysis (WHOSE[2]) that concentrates on the inclusion and application in arbitrary IR systems with different functionality and technology stacks. It allows session-based analysis of user behavior in different systems, in different domains and with different domain knowledge. In WHOSE a whole-session is considered technically as a collection of actions a user performed from starting the system until closing the web browser session. System operators can define actions and filters to meet their individual requirements. All preprocessing, management and presentation of data is then handled by WHOSE. How this can be done is shown in section 4 where we report on experiences we made while applying WHOSE for analyzing log data from Sowiport[3]. WHOSE's graphical user interface consists of an interactive visualization, several filters and detailed session lists. It allows researchers to interactively explore user search behavior based on session data.

## 2    Related Work

The classical IR approach handles the search process as a single-query and multiple documents problem and is for example measured by the TREC evaluation campaign [23]. A more complex scenario arises by the investigation of user sessions. After posing an initial query, users often reformulate their search query until they are satisfied with the results. These multi-query sessions need other evaluation metrics [11]. Furthermore, each search session contains subtasks with explicit cognitive costs (e.g. scanning result lists), which can be addressed using a cost model based on time [1]. Belkin [4] proposes the measure of usefulness for the evaluation of entire information seeking episodes. He distinguishes usefulness in respect to (1) the entire task, (2) each step of interaction and (3) the system′s support for each of these steps.

Longitudinal tasks over several sessions can be identified either by unique user ids or by machine algorithms. Jones et al. [10], for example, identified fine-grained task boundaries in a web search log by using different classifiers and machine learning.

---

[1] http://www.nii.ac.jp/shonan/blog/2012/03/05/whole-session-evaluation-of-interactive-information-retrieval-systems/

[2] Open Source code is available at https://git.gesis.org/public

[3] http://sowiport.gesis.org

Kotov et al. [13] also tried to identify longitudinal tasks which are distributed over several search sessions. They used supervised machine learning with different classifiers to handle identification of cross-session search behavior in web logs. Liao et al. [15] extract task trails from web search logs in contrast to search sessions. They found that user tasks can be mixed up in search logs because of the chronological order and the behavior of users to conduct concurrent tasks in multiple tabs or browsers. Identified tasks seemed to be more precise in determining user satisfaction in web search.

There are different measures and indicators that have been found to be important for session behavior. Fox et al. [7] conducted a user study in web search to find implicit measures that correlated best with sufficiently determining the user satisfaction. It was found that a combination of clickthrough, time spending on the search result page and how a user exited a result or search session correlated best with user satisfaction. Liu et al. [16] conducted a laboratory experiment in which they checked different measures influencing session behavior for different tasks. Three main behavioral measures were identified as important for document usefulness: dwell time on documents, the number of times a page has been visited during a session and the timespan before the first click after an query is issued. Dwell time showed to be most important, however, differs much in cut-off time and needs to be adaptive to different task types. Predictive models has then been applied to the TREC 2011 Session Track and showed improvement over the baseline by using pseudo relevance feedback on the last queries in each session.

The Interactive Probability Ranking Principle (IPRP) [8] is a theoretical framework for interactive information retrieval. It models the search process as transitions between situations. A list of choices is presented to the user in each situation, which can be e.g. a list of query reformulations, related terms or a document ranking. The user decides for one choice and is moved to the next situation. Each choice is connected to the parameters (i) effort, (ii) acceptance probability and (iii) resulting benefit. The overall goal of IPRP is to maximize the expected benefit by optimizing the ranking of choices. IPRP parameters can be derived from observation data like search logs, eye tracking [22] or mouse tracking. Resulting transition models for domains, tasks or subtasks can be visualized with Markov chains. Another popular visualization type that has been used in the field of website analysis for the visualization of user paths is node-link diagrams [6, 17, 24]. Very related to the area of whole-session analysis in IIR is also the field of visual web session log analysis, e.g. for the analysis of website behavior [14] or search usage behavior [19]. One goal of this kind of tools is to identify usage patterns that lead to successful completion of sessions, e.g. to finish a certain task in e-commerce.

There are already a number of frameworks to conduct controlled IIR evaluations. The Lemur Query Log Toolbar[4] is a web browser plug-ins that can capture user search and browse behavior as well as mouse clicks and scrolling events for web search sessions. ezDL [3] is an interactive search and evaluation platform. It supports searching heterogeneous collections of digital libraries or other sources, can be customized and extended, and provides extensive support for search session evaluation including mouse, gaze and eye tracking. Bierig et al. [5] present a framework to design and conduct task-based evaluations in Interactive Information Retrieval. The system focuses on handling multiple inputs from mouse, keyboard and eye tracking. Hall and Toms [9] suggest a common framework for IIR evaluation which also includes components for logging user actions. The task workbench can handle pluggable components like a search box, search results etc. that can communicate with each other. The result is a rich log file where each component contributes detailed information. WiIRE [21] is a web-based system for configuration and conducting IIR experiments which incorporates essential components such as user access, task and questionnaire provision, and data collection. The same idea has been taken by SCAMP [18], a freely available web-based tool for designing and conducting lab-based IIR experiments, which includes all major processes from participant registration to logging and tracking of tasks. The intended benefit of all these frameworks is mainly for controlled IIR experiments in which users conduct several tasks in a laboratory setting. Evaluation data is recorded with logs, mouse, gaze, eye tracking and questionnaires. These controlled data sets can then be used for analysis of search behavior in a single system. However, these toolkits are not intended for the integration into existing IR systems, for the use of uncontrolled log data, their processing and the interactive analysis of user search behavior.

---

[4] http://www.lemurproject.org/toolbar.php

## 3      The Whole-Session Analysis Tool

In this section we present the general functionality of the analysis tool WHOSE: how user interaction data can be logged easily in different environments, how it can be mapped to actions and how data is preprocessed. Finally, we give a general overview of the user interface.

### 3.1    Logging Interaction Data

In IR systems user interactions can be logged in different ways. User interaction data can be recorded anew in various formats with different information depth or may already exist e.g. in form of web server log files.

A common approach is to record user actions in a well-defined schema (e.g. as in [9]). Here, the use of a certain schema has to be fixed and a list of possible interactions with its parameters has to be determined in advance by system experts. Then, the IR system has to trigger a new record to the log if the user applies a certain action. This can be quite a challenge in a real-world IR system if it is proprietary software, closed source or older code, because interceptors need to be implemented at various points in the source code which catch dozens of different user actions.

To overcome this issue, we implemented a logging approach that can handle uncontrolled data either from (1) function calls or (2) from existing log files and later maps them to a structured schema. The benefits here are that the logging component can be very easily implemented into existing software at only one central point in the source code and that existing uncontrolled log files like from the web server or application server can be used for analysis.

In web-based IR systems function calls are often realized by reloading the web page with additional GET/POST-parameters, calling JavaScript or internal AJAX/Servlet or other calls. Function calls contain a string which identifies the action (via function name or parameter) and several additional parameters. For example, in the discovery framework VuFind[5] (used in Sowiport) a simple search can be identified by the URL parameter "lookfor=" followed by a keyword. Similarly other user actions like exporting or adding an item to favorites can be identified. We found that, for example, in VuFind up to 90% of all user interactions can be identified by URL parameters, few interactions are conducted by AJAX or JavaScript calls.

Technically, all function calls can easily be intercepted by some lines of code and can be logged in a database or to a file. Function calls are handled as simple strings and no parsing or extraction is carried out. This makes the logging process very simple and adaptive for the application into many different contexts, be it a different domain, a different technical system or a different functionality. We used this approach in the new version of our IR system Sowiport that has been launched in April 2014. Here, we added an interceptor function at the main class that adds a new entry to the logging table in the database with every reload of the web page. The logging schema for WHOSE only contains very basic fields: "session-id", "user-id", "timestamp", "resultlist_ids", "url" and "referrer-url". "Session-id" is a unique session identifier which is generated in most IR system software. "User-id" is a unique user identifier provided by the IR system. "Resultlist_ids" contains a list of document identifiers from the result list if a search has been conducted. The field "url" contains the string with the requested URL, AJAX or other function calls. The field "referrer-url" contains the URL the systems user requested before the current action.

To test the other approach of handling data from log files, we used an existing database table from an older version of Sowiport with seven years of user data consisting of eleven million rows (with a size of 2GB) and transformed it easily into the necessary table structure.

### 3.2 Mapping Actions

In a next step the logged action data have to be mapped to concrete user actions. Every IR system provides different functionality and the representation in function calls or other uncontrolled data may be implemented differently. Therefore, WHOSE requested a mapping table in which a system expert can specify the mapping between defined user actions and corresponding parameters in the log data. For example, the user action "request search results for search term 'religion' is mapped to the log data entry www.xy.com/results?searchterm=religion". The goal is that the whole logic which is specific for an IR system is collected and defined in this table.

The mapping table is a simple table in CSV format that can be edited in any spreadsheet software. For every action in the IR system (such as searching, filtering or opening the detailed view) the expert needs to define a mapping. Actions are described with an internal and language specific labels and are identified by the system with regular expression patterns. Table 1 shows a row from the mapping table that identifies a simple search action from the homepage. To identify the action the "url" and the "referrer" field from the logging table needs to match to the regular expressions defined in the "url_param" and "referer_param" fields.

---

[5] http://vufind.org/

In addition to the action mappings a group of mappings exist to extract entities like search terms, document ids or result list ids from function calls or strings. So far, we have implemented two operations for entity extraction: (1) *text* means that strings are extracted by the regular expression group functionality, e.g. for extracting query terms; (2) *field* means that the field is directly taken from the logging table into the analysis table, e.g. for logged document ids from the result list.

Table 1. Mapping rule for a simple search

| Referer URL (referer_param) | URL (url_param) | Action |
|---|---|---|
| http:\/\/xy.com\/$ | \/search\/results\? | Simple search from the homepage |

### 3.3 Data Preprocessing

In a preprocessing step WHOSE used the mapping table to transform every row from the logging table into one or several user actions. In addition, further data such as session or action duration are computed and entities like search terms or document ids are extracted. The preprocessing step allows the reduction of data complexity, the mapping to simple actions and the creation of an analysis table. WHOSE can then utilize database functionalities like querying, grouping, indexing and calculation to query subsets and compute additional parameters much faster.

The computational effort for preprocessing can be quite high. Every row from the logging table needs to be matched against all mapping and extraction rules. Here the flexibility of regular expressions results in high computational costs. To improve performance the WHOSE tool uses the Java 6 Concurrency Library to split the work to multiple cores and threads.

### 3.4 User Interface

WHOSE's user interface consists of three parts: (a) filters for time, session and action parameters, (b) an overview visualization and (c) the detailed session list (see Figure 1). In general the design mantra of Shneiderman from the field of Information Visualization is applied: "Overview first, then filter and zoom, details on demand." [20]. This means, users can first get an impression of the overall session dataset with the overview visualization, and can then use filters and time restrictions to filter the data set to specific situations. Filtered user sessions can then be overviewed again in the visualization and in detail in the session list.

The upper part of the user interface contains components to filter the data set by time (Figure 1a). Users can choose from a list of time units (all, last 7 days or 30 days, etc.) or they can set the start and end date explicitly. Directly below, a series of filters are shown which allows the user to filter the whole data set. So far, we have implemented the following set of filters: (1) session contains text (e.g. search terms, facets etc.), (2) session duration, (3) show only sessions of users that are logged in, (4) sessions with a specific user-id, (5) sessions with more than x actions, (6) sessions that contain a certain action, and (7) action duration. Filters can be combined, which means for example that the data set can be filtered for all sessions which contain a certain keyword, and with a document view dwell time over 30 seconds. Additional filters could be implemented easily since the filter functionality relies on SQL-Filtering.

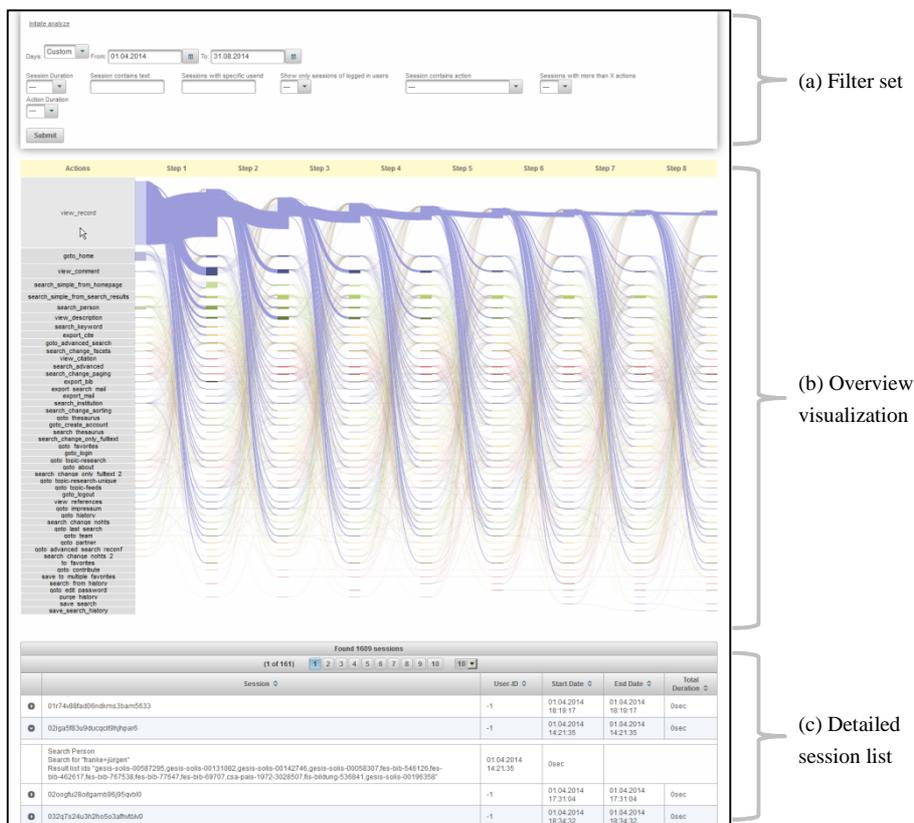

**Fig 1.** Screenshot of the user interface with session data from Sowiport

We chose Sankey diagrams as an overview visualization (Figure 1b) for the actual set of sessions. Each row represents a specific action (e.g. a simple search), each column represents an ongoing search step in the session (first, second, third and so on). Actions are ordered from top to bottom by their highest occurrence within the first eight search steps. The height of the boxes at each search step represents the share of how often this action has been performed in this step. Bézier curves between the box-es show which portion goes to which action in the next step. Hovering with the mouse over an action label highlights the flows for this action and shows which actions have been performed in subsequent steps. The overview visualization in combination with filters can be used to identify user behavior patterns for specific situations.

The session list (Figure 1c) contains all user sessions that fit to the actual time span and filters. Here, the tool user can analyze in detail which actions including their parameters within a session have been performed. Sessions are ordered by descending date and can be unfolded to show all actions within a session.

## 4 Case Study: A First Look into User Behavior in Sowiport

In the following section we present how the tool can be used to analyze a large data set from a real-world IR system. Sowiport is a Digital Library for Social Science information. It contains more than 8 million literature references, 50,000 research projects, 9,000 institutions and 27,000 open access full texts from 18 different databases. Sowiport is available in English and German and reaches about 20,000 unique users per week. The majority of Sowiport's users are German-speaking. The portal has started in 2007 with a major redevelopment in April 2014 based on the VuFind framework and several extensions.

### 4.1 Data Preparation

Every search action in Sowiport is recorded in a logging table with fields like "timestamp", "url", "referrer-url", "result-list-ids" and "user-id". We used data logged between April 2014 and August 2014 consisting of around 2.5 million rows (about 800MB data). A mapping table has been created by system experts which defines about fifty actions and mapping rules specifically for Sowiport. The mapping rules have been tested with regard to completeness and correctness by comparing the system's logging data with screen recording data of six participants who were asked to use Sowiport over a time period of 10 minutes.

### 4.2 Data Analysis

The data analysis starts with a click on the "Submit"-Button that prepares all data for creating the visualization and the session list shown in Figure 1. At the beginning a broad overview of the dataset is provided by showing all log data. The diagram in Figure 1 shows that a large portion of users start their session with the action "view record". These are users that enter Sowiport directly from web search engines, where all detailed views of metadata records are indexed as individual web pages. The four main actions following step 1 can be identified as looking at another record, looking at the comments, looking at the abstract or initiating a new search. This pattern then reoccurs in the following steps.

The data can be filtered by different situations with specific attributes. For example, it can be checked if the search behavior of users that are logged in differs from those who are not. Figure 2 illustrates the results after applying the filter "show only sessions from logged in users". The main entry point for the filtered dataset is the homepage. Then, a large part of users continue with a simple search. In the third step, logged-in users go to a detailed view of a record, conduct another search or restart from the homepage.

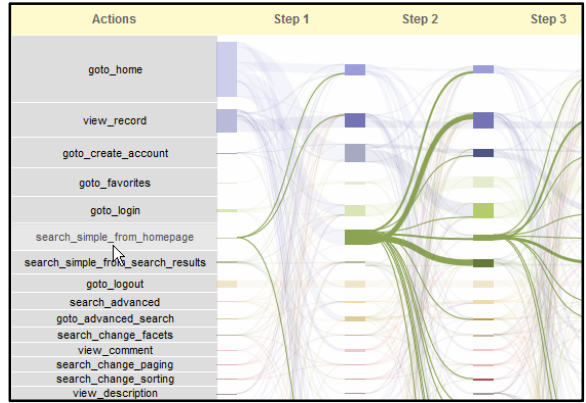

**Fig 2.** The overview diagram shows action patterns for sessions filtered to logged in users with focus on the action "simple search"

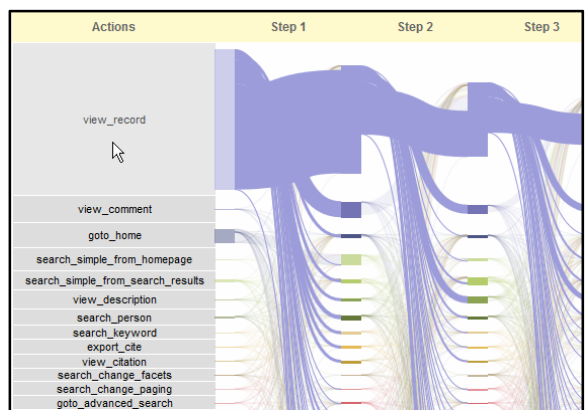

**Fig 3.** Sessions that include a full view of more than 30 seconds

In Figure 3 another example is illustrated. A researcher wants to find out when a session can be considered to be successful. This can be for example, a detailed view of a record, exporting the record or adding the record to a favorite folder. Any of these cases can be described with filters. For example, in our case all sessions are displayed in which records were viewed longer than 30 seconds. The researcher can check which action patterns lead to these situations. Finally, the resulting sessions can be further inspected in the detailed session list. Individual sessions can be opened and all its actions with parameters and durations are shown (see an unfolded session in Figure 1c).

### 4.3 Expert Evaluation

To gain insight on the way the tool can be used, we performed a first user study of WHOSE with two information science lecturers (one female and one male participant) from the Cologne University of Applied Sciences. We decided to use a real-world scenario so that our participants did not have to speculate about the intention of the users and could better concentrate on providing feedback to WHOSE. During a lecture in 2014 their students were assigned to perform a research task with Sowiport to a self-selected topic over a period of four weeks. Our participants used the tool to find out how the students used Sowiport to fulfill their assignments. The objective was not to analyze every single student's behavior but to identify typical search strategies or particularities. The test took about 45 minutes and the participants were asked to use WHOSE and to tell the experimenter everything they noticed, what they considered to be good or problematic, and what would further be needed for improvement.

Their comments give us valuable starting points for the further development of WHOSE. For example, they stated that the Sankey diagram is very complex at first sight and that more interaction opportunities on the diagram would be needed to show and hide selected paths or actions. Furthermore, it is essential for them that the diagram and the detail session list are well connected. Selecting an action in the list for example should trigger highlighting the path within the diagram. Vice versa selecting an action in the diagram should result in updating the table. Also it was observed, that the Sankey diagram helped the participants to identify main paths and to identify actions that are not often used but it did not provide information about absolute action frequencies. The participants suggested that providing a selection of several diagram types would help to be able to assess different aspects in more detail. As the Sankey diagram currently only shows a chronology of search steps, one participant asked for an opportunity to analyze the context of an individual action. She wanted to know which actions have led to a specific action and what the next actions were, independently of the point at which the action has been performed during the search session. On the whole, both participants saw high potential in using a further developed version of WHOSE for tracking typical or individual search steps, for identifying search strategies and furthermore for providing hints for usability problems.

### 5 Conclusion & Future Work

In this paper we introduced WHOSE, a tool for whole session evaluation. The goal is to analyze user search behavior in arbitrary IR systems. The presented mapping concept, based on function calls and user actions, allows not only looking at new recorded log data but also to analyze older log files possibly in different formats and from different systems. This makes it possible to compare or even to aggregate user search behavior of several IR systems in a uniform manner. The graphical user interface allows analyzing sessions of several users at the same time as well as on single user basis. Different filters are provided to reduce the amount of data to specific search situations. Thus, WHOSE can help domain experts and researchers for example to identify situations in which a session is successfully terminated and furthermore which behavioral patterns may lead to these situations. This can be a profound basis to understand at which points in the search process certain difficulties exist and the user can be further supported. Difficulties in the search process can arise from simple usability problems to more complex problems like missing search or domain knowledge. In the sense of the IPRP model [8] the latter problems can be addressed with a list of choices that suggests certain moves in the sense of Bates [2] up to different search strategies and value-added services that supports the user in successfully continuing the search process. In future work, we want to address this problem by automatically identifying critical situations and suggesting supporting services to the user. In addition, we plan to conduct a more comprehensive user study with the next version of WHOSE as well as to perform an expert workshop to identify a first set of typical user search behavior patterns.

**Acknowledgements:** The authors thank our colleagues from the department CSS for a working version of the mapping table for Sowiport.